\documentclass[amsmath,amssymb,prl,twocolumn]{revtex4}
\usepackage{graphicx}

\newcommand{\beq}{\begin{equation}}
\newcommand{\eeq}{\end{equation}}
\newcommand{\beqa}{\begin{eqnarray}}
\newcommand{\eeqa}{\end{eqnarray}}
\newcommand{\non}{\nonumber}
\newcommand{\smax}{\Sigma_\mathrm{max}}
\newcommand{\ssmax}{\Sigma''_\mathrm{max}}
\newcommand{\fmax}{f_\mathrm{max}}

\begin{document}

\title{Numerical study of  metastable states in Ising spin glasses}

\author{Andrea Cavagna, Irene Giardina and Giorgio Parisi}

\affiliation{Dipartimento di Fisica, Universit\`a di Roma ``La Sapienza'' and \\
Center for Statistical Mechanics and Complexity, INFM Roma 1, \\
Piazzale Aldo Moro 2, 00185 Roma,  Italy}

\date{December 19, 2003}

\begin{abstract}
We study numerically the structure of metastable states in the Sherrington-Kirkpatrick spin glass. We 
find that all non-paramagnetic stationary points of the free energy are organized into pairs, consisting in 
a minimum 
and a saddle of order one, which coalesce in the thermodynamic limit. 
Within the annealed approximation, the entropic contribution of these states, that is the complexity,  
is compatible with the supersymmetry-breaking  calculation performed in 
[A.J. Bray and M.A. Moore, {\it J. Phys. C} {\bf 13} L469 (1980)].
This result indicates that the supersymmetry is spontaneously broken in the Sherrington-Kirkpatrick model.
\end{abstract}

\maketitle

In this last year there has been an outburst of new interest in the complexity of disordered systems, 
and in particular of spin-glasses \cite{zec,brst1,brst2,brst3,
leuzzi1,leuzzi2,abm,plefka}. The complexity is the entropic contribution due to 
the exponentially large number of metastable states, and 
in mean-field disordered models it can be computed by calculating the number of minima of the 
Thouless-Anderson-Palmer 
(TAP) free energy \cite{tap}. Metastable states are crucial for the dynamical behaviour of disordered systems, 
and so is the 
complexity. Most notably, in glasses and 
some spin-glasses \cite{glass,pspin} the complexity triggers a dynamical transition which is not 
associated to any static transition. Issues related to the analytic and numerical calculation of the complexity 
are therefore very relevant.

At the origin of the recent new studies there has been the observation done in \cite{brst1}, that the original calculation of the complexity for the Sherrington-Kirkpatrick (SK) model \cite{sk}, performed by Bray and Moore  in 1980 \cite{bm1}, breaks an intrinsic symmetry of the problem, namely a generalized form of the Becchi-Rouet-Stora-Tyutin (BRST) supersymmetry \cite{brs,juanpe}. Moreover, the supersymmetric complexity computed in \cite{brst1} was found to be much smaller that the supersymmetry-breaking complexity of \cite{bm1}. This discovery reopened, after more than twenty years, the problem of how to compute the complexity in the 
SK model, and, more in general, it raised the  question of whether the BRST supersymmetry is 
spontaneously broken or not in disordered systems.

At the theoretical level, the situation is presently rather open. After the annealed calculation of \cite{brst1}, 
the supersymmetric (SS) complexity
has been computed also at the quenched level \cite{brst2,brst3}, showing that it is consistent with the static solution 
of the SK model and that it coincides with the one obtained from the Legendre transform method 
\cite{monasson}, at any level
of replica symmetry breaking. 
However, it has been shown in \cite{leuzzi1,leuzzi2} that the SS complexity  is stable only at the ground state free energy, 
where by definition it is zero. Thus, were the supersymmetry  unbroken, this last result would lead to the conclusion that in the SK model the complexity of metastable states is in fact zero.  On the other hand, the supersymmetry-breaking (SSB) 
complexity (which has only been computed at the annealed level \cite{bm1}) 
is non-zero and stable on a finite range of free energy densities. Moreover, it has recently been proved in 
\cite{abm} that the SSB complexity describes pairs of solutions of the TAP equations which are either minima or 
saddles of order one of the free energy. According to \cite{abm}, the two solutions of each pair coalesce in the 
thermodynamic limit. In conclusion, the SS and SSB predictions differ considerably.

Given the ambiguous theoretical situation, a way to discriminate between the two pictures above, 
and thus to understand whether or not the BRST supersymmetry is spontaneously broken in the SK model, is to perform a numerical experiment. Of course, one may question what would be the dynamical role of the marginally unstable states described by the SSB complexity, but we do not deal with this problem in the present work. Our goal here is just to understand what is the structure of TAP solutions in the SK model, and which theoretical framework correctly describes this structure.

Our numerical results show that:
i) all solutions of the TAP equations are either minima or saddles of order one of the free energy; 
ii) the solutions are organized into minimum-saddle pairs, connected along a mode which is softer the larger the system size; 
iii) the free-energy difference of the paired solutions decreases with increasing system size;
iv) the annealed complexity of TAP solutions as a function of their free energy density is consistent with the SSB complexity of \cite{bm1}. Our results therefore support the idea that the BRST supersymmetry is in fact broken in the SK model, and that the SSB complexity of \cite{bm1} is correct within the annealed approximation.

The TAP equations for the SK model are,
\beq
\tanh^{-1} m_i +\beta^2(1-q)m_i-\beta\sum_j J_{ij} m_j =0 \ ,
\label{tap}
\eeq
where $m_i$ are the local magnetizations, with $i=1,\dots,N$, and the random couplings $J_{ij}$ are
drawn from a Gaussian distribution of mean zero and variance $1/N$. These equations are obtained
by finding the stationary points of the TAP free energy,
\beqa
F_{TAP}(m)=-\frac{1}{2} \sum_{ij} J_{ij} m_i m_j -\frac{N}{\beta} \log 2 -\frac{N\beta}{4}(1-q)^2  \non \\
+ \frac{1}{\beta} \sum_i  \  \log(1-m_i^2)/2  + m_i\,\tanh^{-1}(m_i) \ , 
\non
\label{ftap}
\eeqa
where  $q=(1/N)\sum_i m_i^2$ is the self-overlap of a solution.
We solve numerically
equations (\ref{tap}) by using the C routine BROYDN of Numerical Recipes \cite{bro} starting
with random initial conditions. We analyze
$20429$ samples at $N=20$, $5000$ samples at $N=30$, 
$1000$ samples at $N=40$, $347$ samples at $N=60$ and $77$ samples at $N=80$. 
All data are for $T=0.2 \, T_c$.
In order to make our search of solutions as exhaustive 
as possible, we use two methods. 
First, all solutions we find in a sample are found at least $5$ times. Second, we monitor the Morse sum, 
$\sum_\alpha (-1)^{k_\alpha}=1$, where $k_\alpha$ is the number of negative eigenvalues of solution $\alpha$. 
For $N=20,30,40,60$ we discard samples where this identity is violated for more than $\pm 2$ (the fraction of
discarded samples is always lower than $1\%$). For $N=80$, however, due to the low number of samples, we do not enforce 
this last check. 

\begin{figure}
\includegraphics[clip,width=3 in]{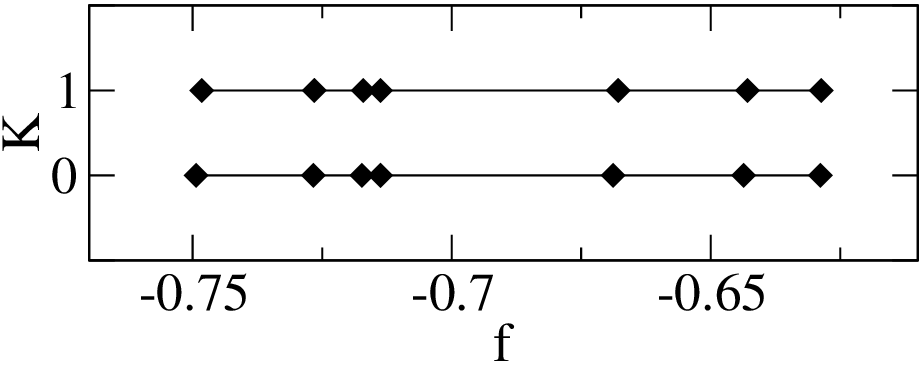}
\caption{Number of negative eigenvalues $K$ vs free energy density $f$ of $14$ non-paramagnetic solutions
of a sample at $N=80$. The solutions are clearly paired into minimum-saddle couples, with very close value of $f$.}
\label{uno}
\end{figure}

The first interesting result is that {\it all} solutions we find, at any value of $N$, are either minima, or saddles with one 
negative eigenvalue of the Hessian (the paramagnet is a minimum). As an example, 
the solutions of a sample at $N=80$ are reported in Fig.1. 
Moreover, all non-paramagnetic solutions are organized into minimum-saddle pairs: by making a gradient descent starting from a 
saddle and moving along the negative mode, we  
reach in one direction a very close minimum \cite{foot}, and in the opposite direction the 
paramagnetic state. Thus, TAP pairs are not directly connected to each other, but they are all connected to the central 
paramagnetic minimum in  a star-like structure. The self-overlap $q$ of a saddle is always smaller than that of the connected 
non-paramagnetic minimum.  The important point is that the two solutions of each pair have very close free energy densities $f$:
by plotting the average free energy density difference $\Delta f$  as a function of $N$, 
we see that $\Delta f$ decreases with $N$ (Fig. 2). A power law fit of the data gives $\Delta f\sim 1/N^{1.26}$.
The paired solutions are connected along the saddle negative mode, which becomes the smallest positive 
mode of the nearby minimum. The fact that the two solutions have very close free energy suggests that this direction is 
extremely flat. In Fig.2  we also  plot the average modulus of the smallest eigenvalue $\lambda_\mathrm{min}$ as a 
function of $N$. A power law fit of the data gives $|\lambda_\mathrm{min}|\sim 1/N^{1.40}$. Thus the direction
connecting the two solutions of a pair is flatter the larger $N$.
The exponent $1/2$ proposed in \cite{abm} seem to fit the data less satisfyingly.

\begin{figure}
\includegraphics[clip,width=3 in]{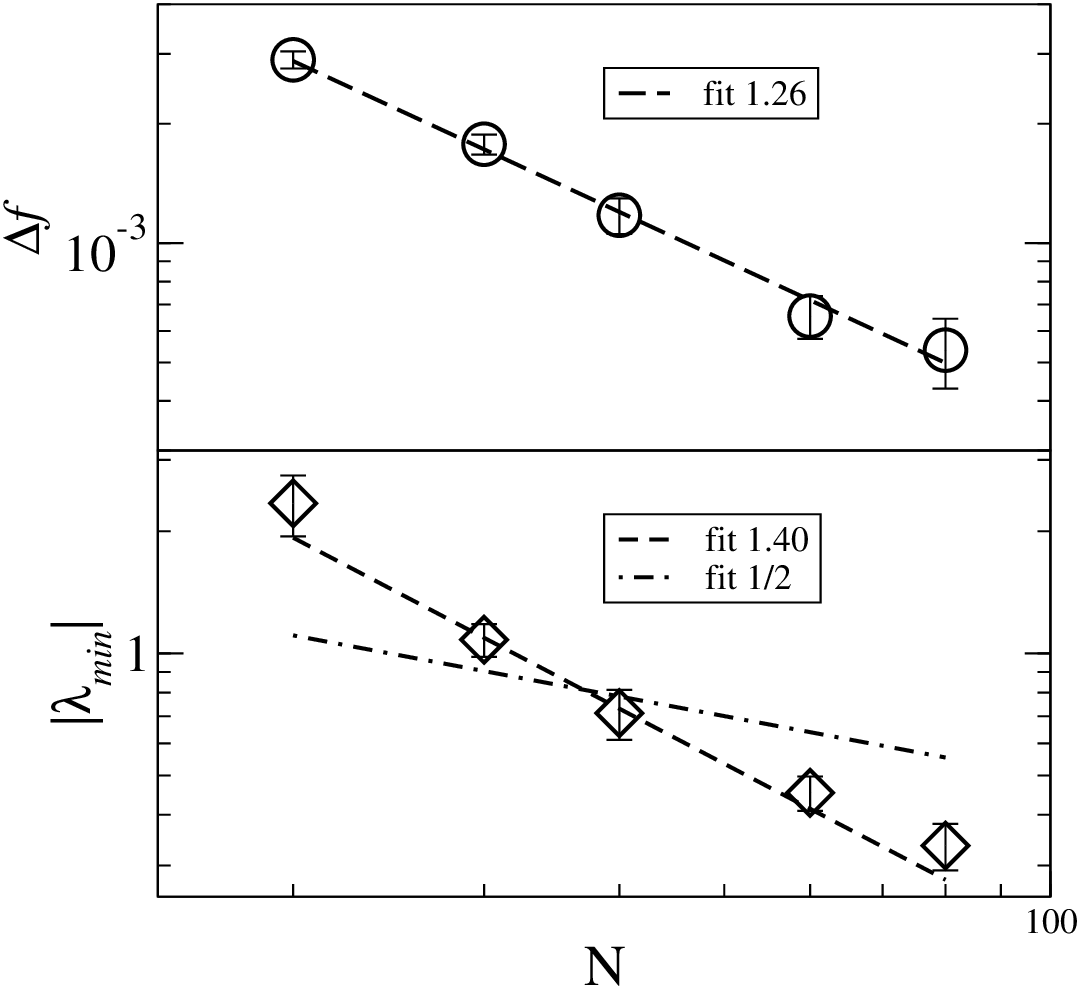}
\caption{Upper panel: the average free energy density barrier.  
Lower panel: the average modulus of the smallest eigenvalue. 
Lines are power law fits (see text).}
\end{figure}

These data suggests that for $N\to\infty$ the two solutions of each pair coalesce, forming a single, 
marginally unstable TAP solution. This conclusion is in qualitative 
agreement with the theoretical results of \cite{abm}, results found in connection with the SSB 
annealed complexity. It is then natural  to ask whether also the annealed 
numerical complexity is compatible with the SSB result 
of \cite{bm1}. The annealed complexity is defined as,
\beq
\Sigma(f) = \frac{1}{N} \log \, {\cal N}(f)  \ ,
\eeq
where ${\cal N}(f)$ is the number of solutions with free energy density $f$, averaged over
the disorder $J$. The SSB annealed complexity is a smooth function of $f$, 
with a maximum at a free energy density $f_\mathrm{max}$. 
Let us define $\Sigma_\mathrm{max}= \Sigma(f_\mathrm{max})$ and $\Sigma''_\mathrm{max}=|\Sigma''(f_\mathrm{max})|$. 
For large $N$ the {\it total} number of TAP solutions $\cal N$
is given by,
\beq
{\cal N}=\int df \ e^{N\Sigma(f)}=e^{N\Sigma_\mathrm{max}} \ 
\sqrt{\frac{2\pi}{N\Sigma''_\mathrm{max}}} \ ,
\label{tot}
\eeq
where we have expanded $\Sigma(f)$ to second order  close to $f_\mathrm{max}$. The probability $p(f)$ of finding a 
solution with free energy density $f$ is therefore given by, 
\beq
p(f)=\frac{{\cal N}(f)}{{\cal N}}= \sqrt{ \frac{N\Sigma''_\mathrm{max}}{2\pi}} \ 
e^{-\frac{1}{2} N\Sigma''_\mathrm{max}(f-f_\mathrm{max})^2 }  \non
\label{pf}
\ ,
\eeq
which becomes a $\delta$-distribution for $N\to\infty$.
For finite $N$ the probability distribution $p(f)$ is a numerically directly accessible quantity. 
In particular, its variance $\sigma$ can be easily checked without the need of any binning of the data.
In Fig.3 we plot the variance $\sigma$ as a function of $N$. 
The full line is the prediction given by the SSB complexity, i.e. $\sigma= 1/\sqrt{N\Sigma''_\mathrm{max}}$, 
with $\smax''^{(\mathrm{SSB})}=8.9$. Numerical data are compatible with the SSB prediction. Moreover, if we perform a fit
to  $\sigma\sim N^{-1/2}$ we find $\ssmax=10.3$, and if we leave the exponent free, fitting to $\sigma\sim N^{-\gamma}$ we get
$\gamma=0.58$ (not shown). We also show the best fit to $\sigma\sim N^{-1/3}$, which is the exponent 
reported in \cite{plefka}. This value, however, does not seem compatible with our data.

\begin{figure}
\includegraphics[clip,width=3 in]{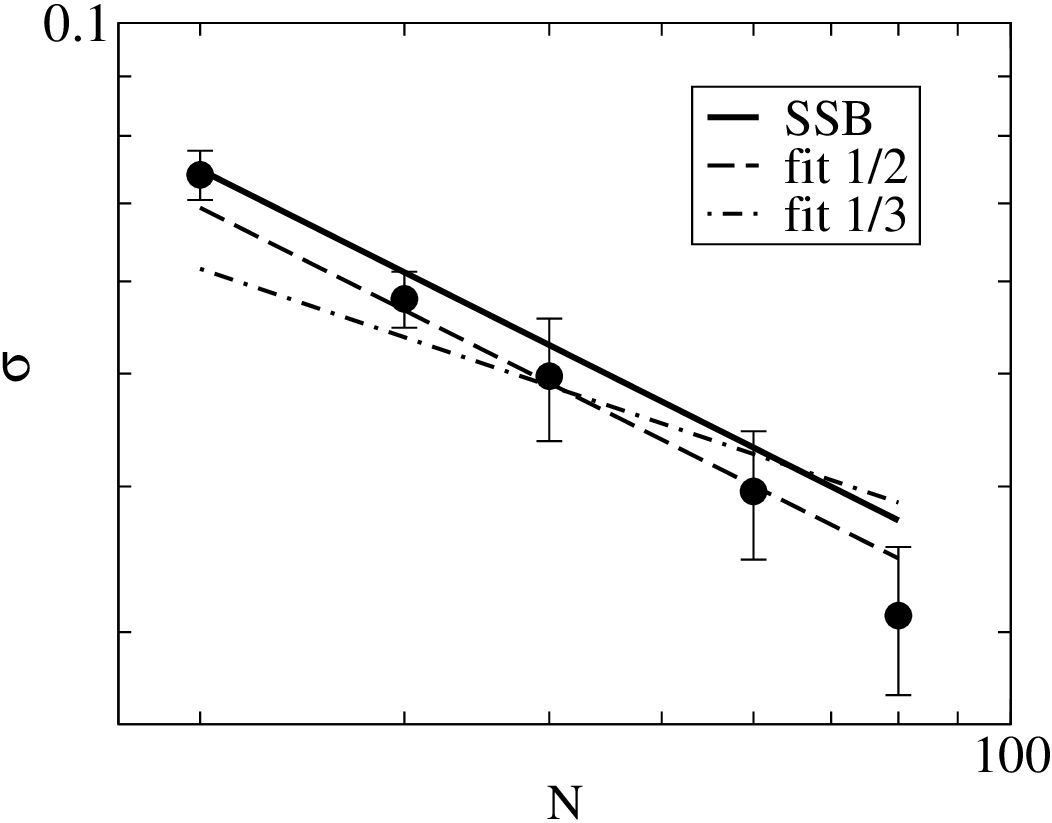}
\caption{The variance $\sigma$ of the TAP solutions, an a function of $N$. Full line: the analytic prediction of the
SSB calculation, $\sigma= 1/\sqrt{N\Sigma''_\mathrm{max}}$. Dashed line: best fit to $N^{-1/2}$. Dash-dotted line: best fit
to $N^{-1/3}$.}
\end{figure}

By making a (arbitrary)  
binning of the data in free energy, we can compute  the full probability $p(f)$ and thus our numerical 
estimate of the complexity,  $\Sigma(f)=\log[{p(f)\cal N}]/N$, where $\cal N$ is the
{\it total} average number of solutions per sample we find numerically. 
Results are shown in Fig.4 and compared to 
the theoretical curve of the SSB complexity $\Sigma(f)$. Even though the numerics is not excellent 
(especially for $N=80$), 
the data seem compatible with the SSB prediction and clearly indicate that the annealed complexity is nonzero. 
However, a more quantitative comparison, not dependent on the binning,  
is clearly desirable. To do this we study the position of the maximum $f_\mathrm{max}$ and the
value of the complexity at this point $\Sigma_\mathrm{max}$, at various $N$. In order to avoid a 
measurement depending on 
the binning, we study the integral of $p(f)$, i.e. $g(f)=\int_0^f dx \, p(x)$, 
and interpolate $g(f)$ 
with a cubic function around its inflection point. In this
way we have an estimate of both $f_\mathrm{max}$ and $\Sigma_\mathrm{max}$, reported in 
Fig.5 as a function of $1/N$. The value of $f_\mathrm{max}$ is basically constant, and it agrees quite well 
with the theoretical SSB value $\fmax^{(\mathrm{SSB})}=-0.654$. 
On the same scale we report the value of the ground state (equilibrium) free energy density 
$f_0$: our data do not seem compatible with the law $\fmax\to f_0$ for $N\to\infty$, proposed in
 \cite{plefka}.
The value of $\smax$ depends more strongly on $N$, as expected from equation (\ref{tot}), i.e.
$\smax^N = \smax^\infty + \log(N\ssmax)/(2N)+\mathrm{O}(1/N)$. The extrapolation for $N\to\infty$ of our data 
for $\smax$ agrees very well with the SSB value $\smax^{(\mathrm{SSB})}=0.052$.

\begin{figure}
\includegraphics[clip,width=3 in]{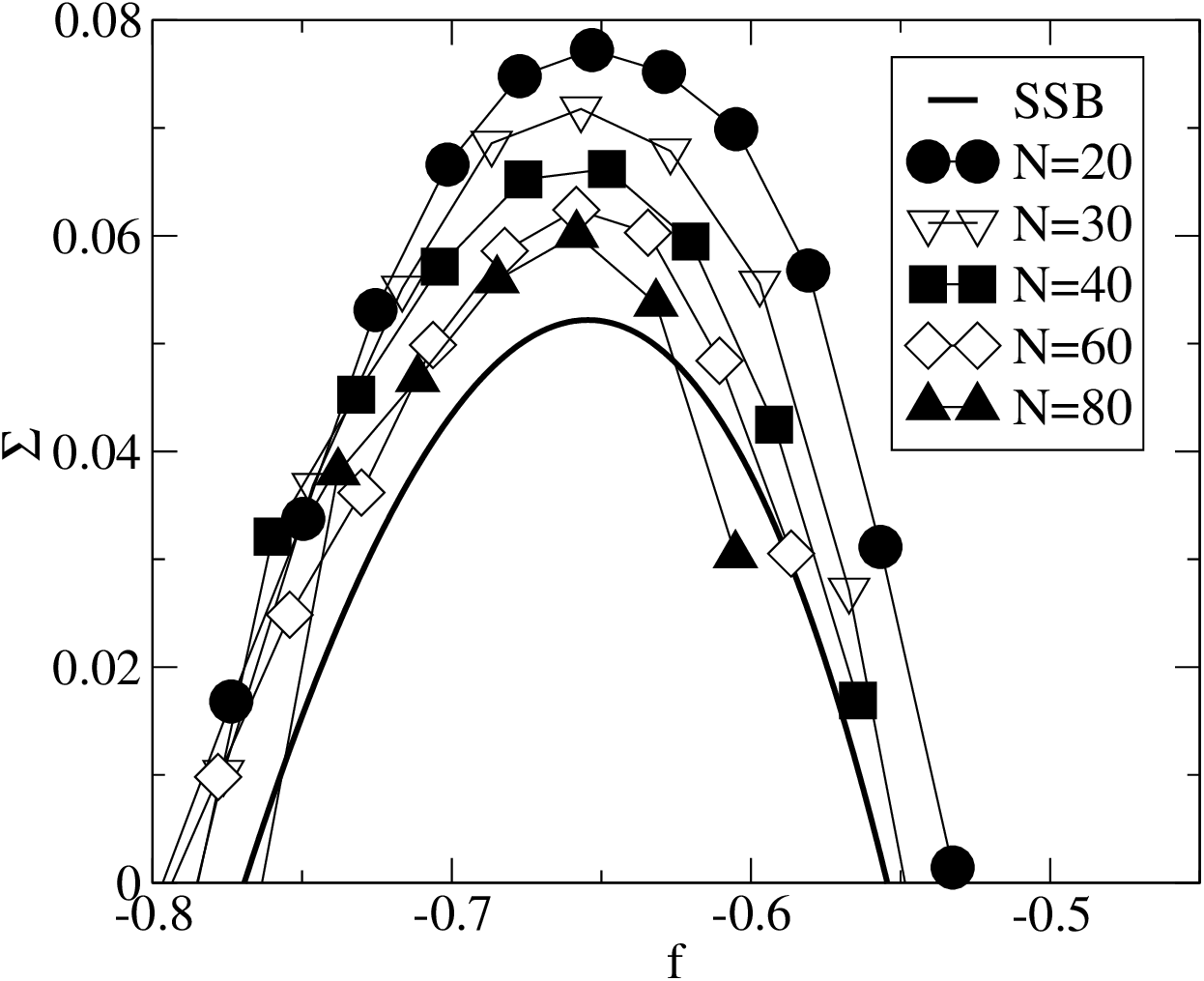}
\caption{Annealed complexity vs free energy density for various $N$; lines connecting the points are only a guide for the eye. 
Full line: analytic SSB prediction.}
\end{figure}

\begin{figure}
\includegraphics[clip,width=3 in]{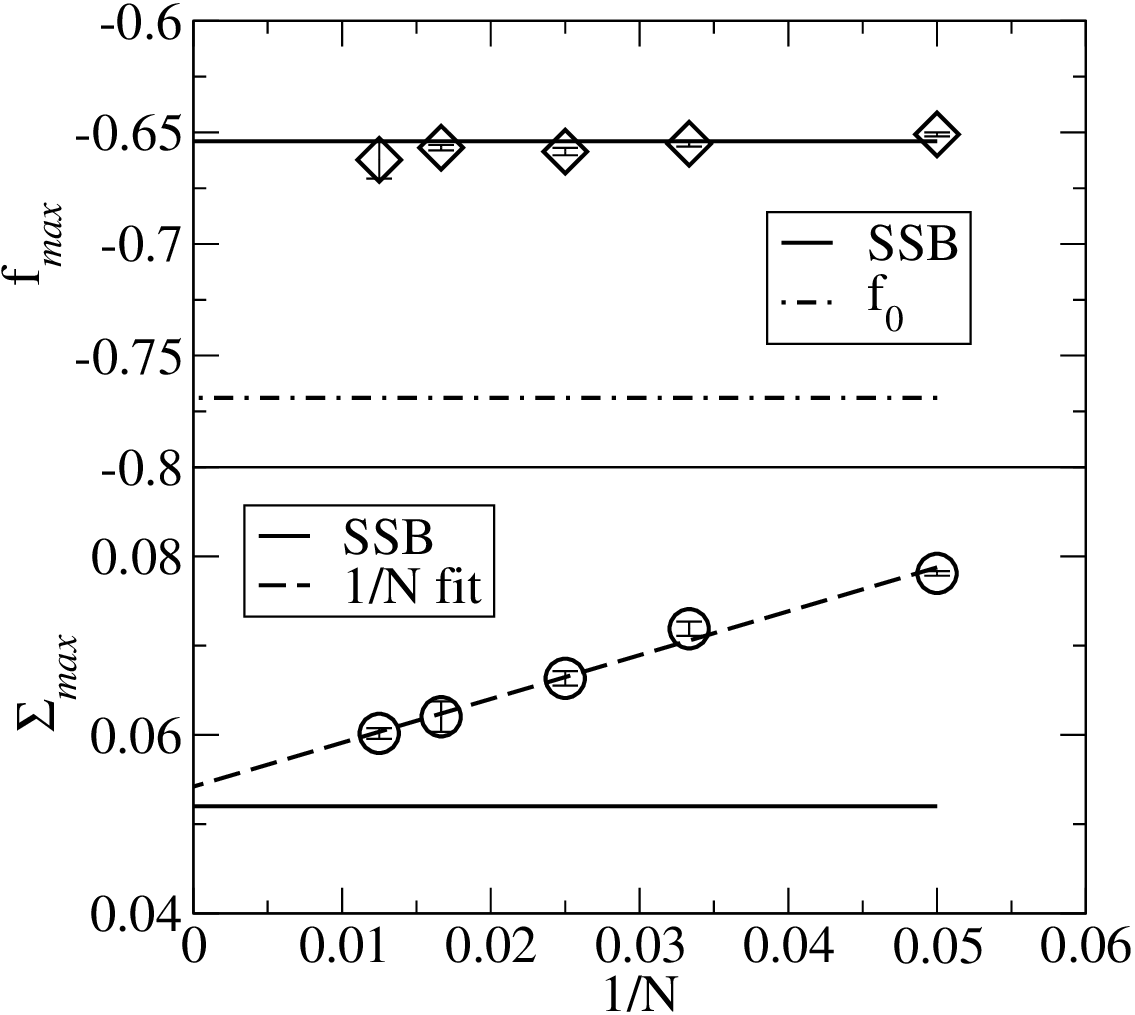}
\caption{Upper panel: $\fmax$ as a function of $1/N$; 
the dash-dotted line is the value of the equilibrium free energy $f_0$. Lower panel: $\smax$, as a function of $1/N$;
the dashed line is a $1/N$ fit to the data. In both panels,
the full line is the $N=\infty$ value of the SSB prediction.}
\end{figure}

A last important consistency check between our numerical data and the SSB calculation concerns the
behaviour of the overlap $q$ and the stability parameter $x_p$, defined as,
$ x_p=1-\beta^2\sum_i(1-m_i^2)^2/N$. In the thermodynamic limit the condition $x_p\geq 0$ must 
be satisfied in order to have a physically
acceptable TAP state \cite{xp1}. In fact, it is precisely this last condition which is violated by the SS 
complexity in all points but the equilibrium free energy $f_0$. Both $q$ and $x_p$ depend on the free energy 
density $f$ of the solutions. In Fig.6 we plot $x_p$ as a function of $q$, parametrically in $f$, at various
values of $N$. We see that $x_p$ is in fact positive for all solutions in the region where $\Sigma(f)\geq 0$.
The data agree quite well with the SSB prediction, even in the phase $f<f_0$, where $\Sigma(f)<0$.

\begin{figure}
\includegraphics[clip,width=3 in]{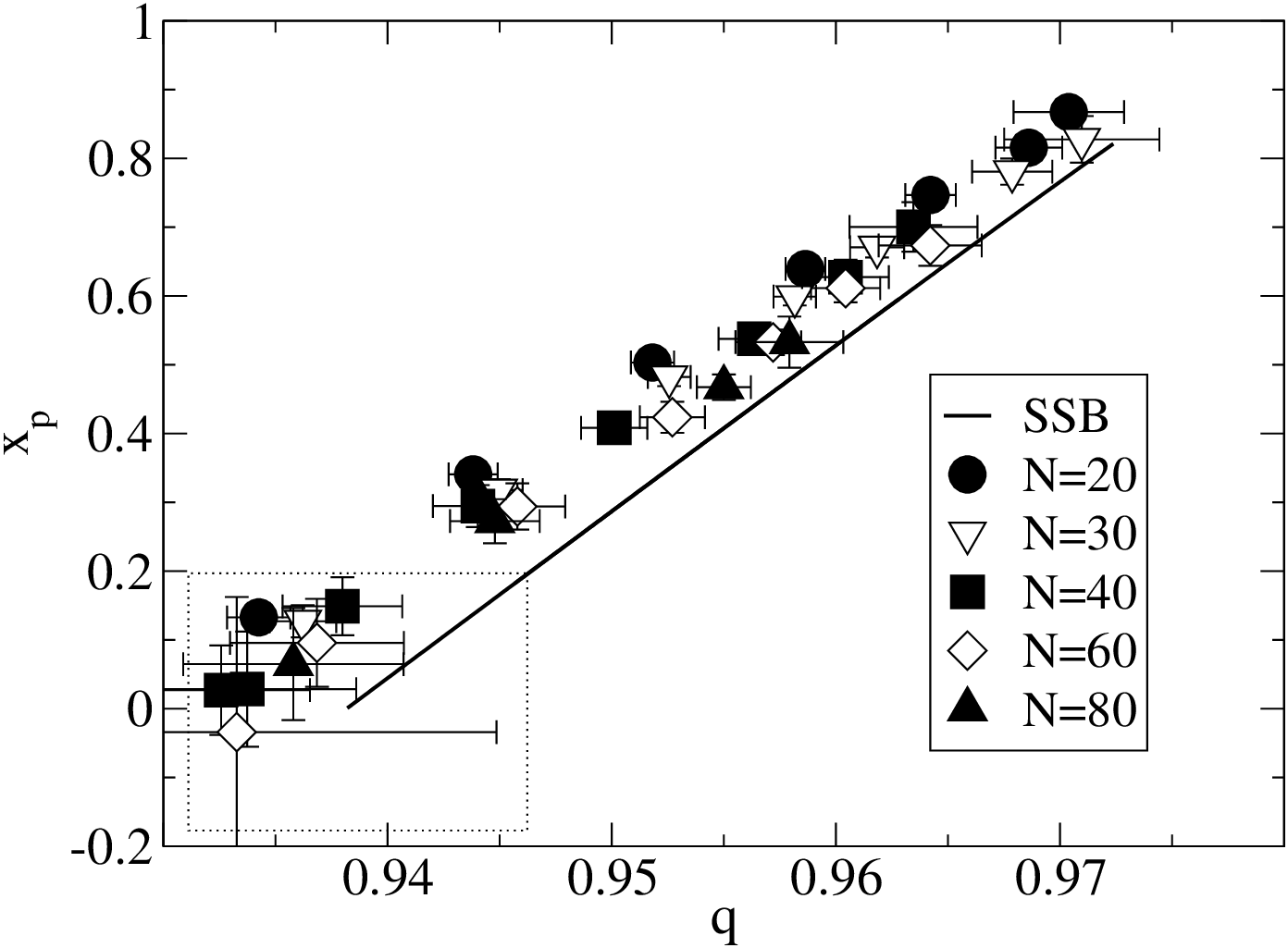}
\caption{The stability parameter $x_p$ vs the self-overlap $q$, parametrically in the free energy density $f$. Full line: the
analytic SSB prediction. Data in the box have $f<f_0$.}
\end{figure}

Our algorithm finds more easily solutions with lower free energy, as can be seen by checking the
frequency each solution is found, and this effect is stronger the
larger $N$.
Therefore, at $N=80$ (where the Morse check is not enforced) we are probably missing a number of
solutions with $f>\fmax$.
We believe this is the reason why the values of $\sigma$ and $\fmax$ at $N=80$ in 
Figs. 3 and 5 are slightly smaller than expected, irrespective of error bars. 
The same problem arises in the reconstruction of $\Sigma(f;N=80)$ for $f>\fmax$ in Fig.4. 
The problem of finding solutions at finite $T$ of the TAP equations is well known, and for this reason 
numerical investigations in the past have been done by using alternative methods: $T=0$ studies of one 
spin-flip stable states \cite{T0}, naive TAP equations \cite{naive}, or modified TAP equations 
\cite{plefkamod}. To the best of our knowledge, the present work is the 
first extensive study of the full TAP equations at finite $T$.

In summary, our results support the conclusion that at the annealed level the complexity of the SK model is 
given by the SSB solution of \cite{bm1}, and therefore that the BRST supersymmetry is spontaneously 
broken in this system. Considering the fact that the supersymmetry is {\it not} broken in the $p$-spin 
spherical  model, it may be argued that systems characterized by full replica symmetry breaking  
(as the SK model), and systems solved by one step of replica symmetry breaking (as the
$p$-spin), belong to two different classes also for what concerns the spontaneous breaking of the 
supersymmetry. The origin of this different behaviour must lie in the geometric structure of  
the states. In particular, the marginality of TAP saddle-minima pairs in the SK model is probably 
the cause of the spontaneous supersymmetry breaking. This fact is also likely to be connected to the different
dynamical behaviour of these two classes of systems: the dynamics of the $p$-spin model is heavily influenced by 
the TAP states, while in the SK model, despite the non-zero SSB complexity, the dynamics asymptotically
reproduces the static results, with no role played by the metastable states. The supersymmetry could then be  
an elegant way of distinguishing different dynamical classes.

We thank A. Bray, L. Leuzzi, M. Moore, T. Plefka, T. Rizzo and the Rome Complex 
Systems Group for discussions, and A. Annibale and E. Trevigne 
for the SSB data. We acknowledge support of the ESF-SPHINX program.


\begin{thebibliography}{22}
\bibitem{foot} The average distance is $d=[\sum_i (m_i^{(0)}-m_i^{(1)})^2/N]^{1/2}\sim 0.1$, weakly dependent on $N$.
\end{thebibliography}
\end{document}